# Temperature effects on the structure and mechanical properties of vapor deposited a-SiO$_2$

V. Jambur [a], M. Molina-Ruiz [b], T. Dauer [b], D. Horton-Bailey [b], R. Vallery [c], D. Gidley [d], T. H. Metcalf [e], X. Liu [e], F. Hellman [b], I. Szlufarska [a, *]

[a] *Department of Materials Science and Engineering, University of Wisconsin, Madison, WI 53706, USA*

[b] *Department of Physics, University of California, Berkeley, CA 94720, USA*

[c] *Department of Physics, Grand Valley State University, Allendale, MI 49401, USA*

[d] *Department of Physics, University of Michigan, Ann Arbor, MI 48109, USA*

[e] *Naval Research Laboratory, Washington, DC 20375, USA*

*Corresponding author:

Prof. Izabela Szlufarska

Department of Materials Science & Engineering

University of Wisconsin – Madison

Madison, WI 53706, USA

Phone: +1 (608) 265-5878

E-mail: szlufarska@wisc.edu

**Abstract**

Amorphous silica (a-SiO$_2$) exhibits unique thermo-mechanical behaviors that set it apart from other glasses. However, there is still limited understanding of how this mechanical behavior is related to the atomic structure and to the preparation conditions of a-SiO$_2$. Here, we used electron beam (e-beam) physical vapor deposition (PVD) to prepare a series of a-SiO$_2$ films grown at different substrate temperatures and then combined molecular simulations with Positronium Annihilation Lifetime Spectroscopy and nanoindentation experiments to establish relations among processing, structure, and mechanical response of the films. Specifically, we found that increase in the growth temperature leads to increase in the elastic moduli and hardness of the films. The relative porosity in the films also increases while the a-SiO$_2$ network itself becomes denser, resulting in an overall increase in density despite increased porosity. In addition, we found that the a-SiO$_2$ films exhibit the same anomalous temperature dependence of elastic modulus as bulk a-SiO$_2$. However, the rate of increase in the elastic modulus with the measurement temperature was found to depend on the density of the a-SiO$_2$ network and therefore on the growth temperature. Our findings provide new insights into the influence of the atomic network structure on the anomalous thermomechanical behavior of a-SiO$_2$ and in turn guidance to control the mechanical properties of a-SiO$_2$ films.

**Keywords**: a-SiO$_2$, vapor deposition, mechanical properties, nanoindentation, PALS, molecular dynamics

1. Introduction

Amorphous $SiO_2$ is known to exhibit anomalous thermo-mechanical behaviors; its elastic modulus increases with increasing temperature [1], the bulk modulus goes through a minimum at around 2 GPa during compression [2], and it has a negative thermal expansion coefficient at low temperatures in the range of 140 K to 210 K [3]. Molecular dynamics (MD) studies have linked the anomalous behaviors to local, reversible polyamorphic transitions in the network structure of a-$SiO_2$ [4]. These transitions involve changes in the medium range order and affect the ring geometry in the a-$SiO_2$ network while the short-range structure (bond lengths, bond angles, and coordination numbers) remains unchanged. However, there is currently lack of experimental evidence for these local reversible transitions due to the difficulty in characterizing the topological changes in the medium range structure of a-$SiO_2$. In addition to the aforementioned anomalies in the elastic response, a-$SiO_2$ has been found to undergo irreversible densification when subjected to high pressure [5]. Experimental [6,7] and theoretical [5,8] studies on this pressure induced transformation have shown that the densification of the a-$SiO_2$ network structure is accompanied by the formation of smaller network rings.

a-$SiO_2$ films and coatings have found key applications in many industries including semiconductors [9], optical systems [10], and photovoltaics [11] to name a few. While sol-gel synthesis [12] and chemical vapor deposition (CVD) [13] are the most commonly used methods to synthesize a-$SiO_2$ films, physical vapor deposition (PVD) is an environmentally friendly alternative that does not require dangerous reactants or high temperatures [14]. In addition, the enhanced surface mobility achieved in some vapor deposited glasses allows those systems to easily explore the energy landscape and find more stable configurations [15] that cannot be accessed through conventional melt-quench techniques that are used to synthesize bulk glasses.

This means that the process parameters in PVD can be tuned to obtain glasses with optimized properties. For any application, it is generally desirable to have films with higher elastic modulus and hardness in order to provide mechanical integrity and to ensure durability. In the case of organic and metallic glasses, PVD process parameters like deposition rate and substrate temperature have been tuned to yield denser films with high elastic modulus and hardness [16,17]. However, there is still a lack of a systematic understanding of the influence of PVD process parameters on the structure and mechanical properties of a-$SiO_2$ films.

In this work, e-beam evaporation, a PVD technique, was used to prepare a-$SiO_2$ films over a range of temperatures from 60 °C to 900 °C. In doing so, we were able to sample a wide range of a-$SiO_2$ structures and study their influence on the mechanical behavior. We also obtained new insights on the thermally induced structural transitions and anomalous thermomechanical behaviors in a-$SiO_2$. Structural changes in terms of the density and porosity were measured using characterization techniques such as Rutherford backscattering spectrometry (RBS), profilometry, and positron annihilation lifetime spectroscopy (PALS), while mechanical characterization was carried out using nanoindentation and double-paddle oscillator. The experimental observations were then interpreted with the help of MD simulations.

## 2. Methods

### 2.1 Experimental

#### 2.1.1 Preparation of a-$SiO_2$ films

a-$SiO_2$ films were prepared by e-beam evaporation of 99.999% $SiO_2$ pieces. The base pressure was ~1×$10^{-8}$ Torr, films were grown at 0.5 Å/s, with substrate temperatures ranging from 60 °C to 900 °C. Some films were capped with a thin (5 nm) Al layer to prevent water absorption [18]. Films thickness was measured by a KLA Tencor ASIQ profilometer, ranging from 800 nm to 1000 nm for nanoindentation measurements, and from 200 nm to 600 nm for ion beam analysis (IBS) and PALS measurements.

### 2.1.2 Ion beam analysis

IBS was performed in the films by means of RBS and elastic recoil detection analysis (ERDA) in a NEC model 5SDH Pelletron tandem accelerator, with a beam energy of 3040 keV. RBS was used to determine the film's stoichiometry, and in combination with profilometry, the film's density. The density is calculated as the ratio between surface aerial atomic density (obtained by RBS) and the film thickness (obtained by profilometry). ERDA was used to quantify hydrogen presence in the films associated with water absorption.

### 2.1.3 Positron annihilation lifetime spectroscopy

PALS is a well-established technique for probing free volume voids in insulating materials [19,20]. In general, positrons from a radioactive source are stopped in the target material where a fraction capture an electron and form the long-lived triplet state of positronium (the hydrogen-like bound state of the electron and its antiparticle, the positron). Positronium preferentially traps in free-volume voids with the measured annihilation lifetime being related to the average void size. In this experiment, with films ranging from 200 nm to 600 nm in thickness, we employed an electrostatically focused beam of mono-energetic positrons in high vacuum [21] to restrict implantation to the target film and to avoid implantation in the supporting substrate. A positron beam implantation energy of 3.2 keV was determined to be optimal for these films.

The beam spectrometer uses a fast-timing system with a time resolution of 0.4 ns to measure the time difference between the implementation of positrons into the film and the eventual annihilation of the positronium. Data is acquired over one day with approximately 3 million events in the time histogram. The histogram is then fit to either a discrete set of positron and positronium lifetimes (and corresponding fractional intensities) [22] or a continuum distribution of lifetimes using the Laplace inversion program CONTIN [23]. Finally, using well-established methods [24], we convert the fitted lifetime continuum distribution into a pore diameter distribution for simple pore model shapes.

### 2.1.4 Nanoindentation

Nanoindentation experiments were carried out using a Hysitron TI-950 Triboindenter with a Berkovich probe. Films grown at 60 °C, 600 °C, 800 °C, and bulk a-SiO$_2$ sample were indented to peak loads in the range 300 µN to 400 µN at a constant loading rate of 200 µN/s. The peak loads were selected such that the maximum depths of the indents were less than 10% of the film thickness to avoid substrate effects. The samples were indented in an N$_2$ atmosphere at temperatures of 25 °C, 100 °C, 200 °C, 300 °C, and 400 °C to study the temperature dependance of their mechanical properties. After cooling back down to 25 °C, the samples were indented again to check if the changes in mechanical properties with increasing temperature were reversible. 15 indents were performed at each temperature. A Hysitron xSol stage was used to control the temperature of the samples during indentation. The hardness H and the reduced elastic modulus E (related to the Young's modulus by Eq. 1) were calculated from the load vs. displacement data using the Oliver-Pharr method [25]. From here on, reduced elastic modulus E will be referred to as elastic modulus.

$$\frac{1}{E} = \frac{1-v_i^2}{E_i} - \frac{1-v_s^2}{E_s} \quad (1)$$

$E_i$ = Young's modulus of indenter

$v_i$ = Poisson's ratio of indenter

$E_S$ = Young's modulus of a-SiO$_2$

$v_S$ = Poisson's ratio of a-SiO$_2$

### 2.1.5 Double-paddle oscillator

Double-paddle oscillators (DPOs), made of 300 μm thick (100) high resistivity silicon, were used to measure the rigidity modulus or shear modulus G of 300 nm thick a-SiO$_2$ films grown at different temperatures. DPOs have an exceptionally small damping of the second antisymmetric mode that, operating at a resonant frequency of ~5500 Hz, yield very high quality factors at low temperature (Q ~ 10$^8$) [26]. Films were measured from 300 mK to room-temperature, and G was determined at room-temperature from the shift on the DPO resonance frequency after the film was deposited on [27].

## 2.2 Simulations

MD simulations were performed using the LAMMPS simulation package [28]. a-SiO$_2$ structures with varying densities were obtained by melting β-cristobalite at 8000 K and quenching to 300 K with a range of quench rates from 0.385 K/ps to 7700 K/ps. The details of the melt-quench procedure can be found in Ref.[29]. The interatomic forces were calculated using a Vashishta potential [30] which has been used extensively for studies of structure and mechanical properties of a-SiO$_2$ [29–33]. The evolution of the structure with temperature and the influence of ring size distributions on the thermally induced local reversible transitions in a-SiO$_2$ was then investigated by heating the melt-quenched structures in the constant pressure-constant temperature (NPT) ensemble.

## 3. Results and discussion

### 3.1 Effect of growth temperature on density and structure

Figure 1 shows the measured densities of a-SiO$_2$ films as a function of growth temperature (T$_{Gr}$). The density increases with increasing T$_{Gr}$, from 89% to 97% of the bulk a-SiO$_2$ density prepared by the conventional melt-quench method. ERDA measurements show that the H concentration in these films, due to water absorption, reduces from 6.2 at. % H to 1.7 at. % H for films grown at 60 °C and 900 °C, respectively. This reduction is attributed to a lower diffusivity of water molecules in the a-SiO$_2$ network as it becomes denser. One of the factors that affects the density of a-SiO$_2$ is the medium-range order, and it is usually characterized by the size distribution of rings in the network structure [34]. These rings are largely invisible to most experimental characterization tools, which makes it challenging to determine the ring statistics [35,36]. The other factor that influences the density of a-SiO$_2$ is porosity due to the presence of voids, particularly nanovoids, in the structure [37]. The sizes of voids and relative porosities in the a-SiO$_2$ films can be compared using PALS which is an excellent tool to probe the free volume in the form of nanovoids in materials.

Positronium lifetimes obtained by continuum fitting of PALS spectra using the program CONTIN are shown in Figure 2a. These distributions can be thought of as the normalized fraction of positronium decays that occur with lifetime τ between τ and τ+dτ. The area under the red curve for the a-SiO$_2$ film with T$_{Gr}$ of 900 °C is arbitrarily taken to be unity and for sample comparison purposes the areas under the green (T$_{Gr}$=600 °C) and blue (T$_{Gr}$=60 °C) curves have been adjusted down from unity by the ratio of each spectrum's total positronium (Ps) formation intensity to that of the 900 °C grown sample. These ratios are 0.89 and 0.81, respectively. The area under each curve is thus a relative measure of the number of positronium annihilations in each spectrum which, together with positronium lifetime (related to pore size), can be used to estimate the relative porosity and pore size distribution for each sample.

Once we assume a pore shape model (in this case spherical pores for simplicity) we can convert each lifetime distribution into a corresponding pore size distribution (PSD). Using the techniques developed in Ref. [24], we have modified CONTIN to fit the lifetime histogram directly into a pore diameter distribution as shown in Figure 2b. The fitting program includes the lifetime to pore diameter conversion plus one further refinement in that it incorporates and compensates for a pore size dependence in the trapping probability of Ps. Since Ps diffuses into pores through the pore surface, larger pores should have a higher Ps trapping probability than smaller pores and therefore larger pores, which have longer lifetimes, are over-represented in the lifetime distribution. The PSD fits shown in Figure 2b assume a trapping probability that is proportional to the surface area of a pore, which partially compensates for the increased volume with diameter in terms of relative porosity contribution. The three PSD's in Figure 2b are plausible fits to a spherical pore diameter distribution. As with the lifetime distributions, the area under the $T_{Gr}$=900 °C spectrum (nominally the porosity) is normalized to unity and the other PSDs are appropriately normalized (0.76 and 0.79 for $T_{Gr}$=60 °C and $T_{Gr}$=600 °C, respectively) as a good approximation for comparison of *relative* porosity. Even though it is challenging to determine *absolute* porosity in PALS, the technique is very useful for relative porosity comparisons of chemically similar materials such as the a-$SiO_2$ films in this study.

As presented in Figure 2b, the spherical pore diameter ranges from about 0.2 nm to 2.0 nm. One of the limitations of CONTIN is that, technically, we cannot definitively distinguish whether there are 3 sharply defined void sizes or whether the film really has a broad continuum, however, in amorphous materials we would favor the latter interpretation. All three samples clearly have a dominant pore diameter D near 1.0 nm and well-resolved large diameter pores near 1.8 nm that, despite the larger volume, contribute less than 7% to the overall film porosity.

The trend of increasing positronium lifetime with increasing $T_{Gr}$ is clear in Figure 2a. Although the trend in the pore diameter with respect to $T_{Gr}$ in Figure 2b is not as clear, based on the normalized areas under the pore diameter distribution, there is still a trend of increasing pore volume fraction (or higher relative porosity) with increasing $T_{Gr}$. Specifically, the normalized areas are 0.76, 0.79, and 1.0 for $T_{Gr}$ of 60 °C, 600 °C, and 900 °C, respectively. These trends both indicate that higher $T_{Gr}$ leads to a higher porosity. The one caveat to this conclusion is that the fitted positronium intensity is more open to alternative interpretation than is the lifetime to pore size conversion. Positronium intensity can be strongly affected by film chemistry, which is independent of porosity, but in this case the films are chemically equivalent regardless of the different amounts of absorbed water found in films grown at different temperatures. For this reason, an equivalent set of Al capped films without absorbed water [18] was also measured with PALS. We obtained the same trends, which demonstrates that absorbed water does not alter the annihilation of Ps within the a-$SiO_2$ films. Therefore, it is reasonable to conclude that pore size and total pore volume increase with increasing growth temperature in PVD a-$SiO_2$ films.

The results from PALS may seem counterintuitive at first since one would expect lower porosity in the films with higher $T_{Gr}$ due to the higher density measured in equivalent films. However, these results are similar to those reported by Ono *et al.*[38] in a PALS study on bulk a-$SiO_2$ glasses with varying densities associated with different quench rates. The authors found larger voids and increased volume of empty space in samples with a higher density. In a different study on the Raman spectra of bulk a-$SiO_2$, samples with higher densities, were found to have a higher fraction of smaller 3- and 4-membered network rings [39]. Based on this Raman study, Ono *et al.*[38] concluded that the presence of a denser network with a higher fraction of small 3- and 4-membered network rings compensated for the larger voids and resulted in a higher density in the samples. Similarly, we conclude that the a-SiO$_2$ network in our films becomes denser as T$_{Gr}$ increases, which explains the increase in density, in spite of the (counterintuitive) increasing porosity. That means, the films with higher T$_{Gr}$ have higher densities and a higher fraction of smaller rings in their network, despite having higher porosity.

### 3.2 Effect of growth temperature on mechanical properties

Elastic modulus E, shear modulus G, and hardness H of the a-SiO$_2$ films measured using nanoindentation and DPOs are shown in Figure 3a and Figure 3b. We observe that both moduli and the hardness of the a-SiO$_2$ films increase with increasing T$_{Gr}$ but remain lower than that of bulk a-SiO$_2$. In order to explain this trend, we first consider the effect of porosity. It has been shown in spin coated silica films that the Young's modulus decreases with increasing porosity [40]. However, we see the opposite behavior in our samples. That is, the films with a higher T$_{Gr}$ have higher moduli despite having higher porosity. This result can be explained when we consider that the density of the films also increases with T$_{Gr}$, and the a-SiO$_2$ network becomes denser. The compactly packed rings in the denser network are more constrained and are expected to be more difficult to deform as compared to the larger network rings present in the films grown at lower temperatures, yielding higher elastic and shear moduli and hardness.

### 3.3 Effect of indentation temperature on mechanical properties

The elastic modulus of the a-SiO$_2$ films as well as the bulk a-SiO$_2$ sample as a function of the indentation temperature (T$_{In}$) is plotted in Figure 4a. Here, we find that the elastic modulus increases with increasing T$_{In}$ in both the films and the bulk a-SiO$_2$. This kind of behavior is typical of a-SiO$_2$ and has been attributed to local, reversible transitions that take place in the silica network with increasing temperature[4]. These transitions resemble the alpha to beta transformation in cristobalite and involve rotation of Si-O-Si bonds about the Si-Si axis. The reversible transitions cause changes in the shape of the rings in the network structure of silica. The soft, asymmetric rings at lower temperatures transform into rigid, symmetric rings after the rotation of the Si-O-Si bonds. As a result, the ability to pivot about the Si-O bond is eliminated and subsequent elastic strains can only be accommodated by bending or stretching of bonds. These processes lead to an increase in the elastic modulus. Indentation of the samples after they had cooled down to room temperature showed that the elastic modulus returned to its initial value (Figure 5), confirming that the structural changes with increasing temperature were indeed reversible.

In contrast to the elastic modulus, hardness decreases with increasing $T_{In}$ for all the samples considered in this study (Figure 4b). This result is rather surprising since the network structure of silica becomes stiffer with increasing temperature due to the local reversible transitions. The reason for this trend in hardness is that hardness is defined as the resistance to indentation and depends on both elastic and plastic deformation in the material. While the changes in ring geometry affect the elastic response of a-$SiO_2$, the plastic deformation is controlled by other factors. Specifically, during indentation, the plastic deformation in a-$SiO_2$ is dominated by densification [41]. This is in essence a displacive transformation involving the breaking and formation of new Si-O bonds, where the network structure collapses and becomes denser. The higher temperatures that induce local reversible transitions that make the a-$SiO_2$ network rings stiffer, also accelerate re-bonding and promote densification. The fact that the hardness is decreasing with $T_{In}$ tells us that the accelerated densification, which causes increased plastic deformation, has a larger impact on the overall deformation of the material compared to the local reversible transitions that increase resistance to elastic deformation.

Next, we look at how $T_{Gr}$ affects the *rate of change* in the elastic modulus and in hardness with respect to $T_{In}$. In Figure 6a, we see that when $T_{In}$ changes from 25 °C to 400 °C, the resulting change in the elastic modulus is the largest for the film grown with the lowest $T_{Gr}$ and it decreases as $T_{Gr}$ increases. The bulk a-SiO$_2$ sample shows the smallest increase in the elastic modulus. As explained earlier, the increase in the elastic modulus with the indentation temperature can be attributed to the local reversible transitions that lead to changes in the ring geometry. These transitions resemble the alpha to beta transformation in cristobalite silica and involve the rotation of Si-O-Si bonds in the silica rings. We hypothesize that the rotation of the Si-O-Si bonds becomes more constrained as the silica network becomes more compact and this results in fewer transitions in the structure and therefore a smaller increase in elastic modulus.

To verify our hypothesis, we analyzed thermally induced evolution of a-SiO$_2$ structures with different ring size distributions using MD simulations. We started with simulated quenching of molten SiO$_2$ over a range of quench rates from 0.385 K/ps to 7700 K/ps. We found that the potential energy per atom increased with the quench rate, resulting in structures with a lower stability (Figure 7a). However, the density of the melt quenched structures did not seem to show any dependance on the quench rate in the regime between 0.385 K/ps and 77 K/ps (Figure 7b). In fact, the densities in this regime were quite close to the experimentally measured density of bulk a-SiO$_2$. However, at quench rates beyond 77 K/ps, the density began to decrease and at 7700 K/ps, it became comparable to the experimentally measured density of the films grown at 600 °C and 800 °C. We then used the RINGS code [42] to compare the ring size distributions in the structures quenched at 77 K/ps and 7700 K/ps and found that the structure that had a higher density had fewer rings with 8 members or more (Figure 8a). These two structures are therefore suitable to test out our hypothesis regarding the effect of the network ring sizes on the thermally induced local reversible transitions in a-SiO$_2$. To study this effect, we heated the two melt-quenched structures up to a temperature of 800 °C and we tracked the rotation of the Si-O-Si bonds about the Si-Si axis using the plane normal correlation defined in Ref.[4]. The plane normal correlation defined as the dot product between the normal to the plane of the Si-O-Si bond at a reference temperature and any other temperature. For example, if we take the reference temperature to be 25 °C, the plane normal correlation at 25 °C will be equal to 1. As the temperature is increased and the Si-O-Si bonds start to rotate, the plane normal correlation will start decreasing. The changes in the plane normal correlation, while going from 25 °C to 800 °C for the structures quenched at 77 K/ps and 7700 K/ps, are shown in Figure 8b. We can see that the plane normal correlation drops at a much faster rate with increasing temperature in the case of the 7700 K/ps quenched structure than for the sample quenched at 77 K/ps. That is, the structure which has a lower density and larger network rings has a higher propensity to undergo the local reversible transitions involving the rotation of Si-O-Si bonds. As a result, such a structure is bound to exhibit a larger increase in elastic and shear moduli. This finding from MD simulations confirms our initial hypothesis that the rotation of the Si-O-Si bonds becomes more constrained as the silica network becomes denser and this results in fewer transitions in the structure and therefore a smaller increase in the elastic modulus.

When we look at how $T_{Gr}$ affects the rate of the change in hardness with respect to $T_{In}$ in our experiments (Figure 6b), we can see that the change in hardness in going from $T_{In}$=25 °C to $T_{In}$=400 °C is the smallest for the film grown with the lowest $T_{Gr}$ and it increases as $T_{Gr}$ increases. The bulk a-SiO$_2$ sample shows the largest drop in hardness. This is the opposite trend to what we found for the elastic modulus (Figure 6a). To explain this trend in hardness, we need to go back to the discussion of the competing effects of densification and the local reversible transitions on the deformation of a-SiO$_2$. The thermally activated densification process accelerates plastic deformation while the local reversible transitions that is most pronounced at higher temperatures result in stiffer network rings, which are more resistant to elastic deformation. When discussing the decrease of hardness with an increasing nanoindentation temperature $T_{In}$ (Figure 5), we concluded that the effect of densification on the overall deformation of the samples was larger than the effect of local reversible transitions. However, we have just shown through MD simulations (Figure 8a and 8b) how the local reversible transitions take place to varying degrees in the glasses depending on their density and ring size distribution. This means that the propensity for undergoing these reversible transitions will be higher in the a-SiO$_2$ films grown at lower temperatures due to the larger network rings. This in turn implies that the increase in stiffness of the network rings due to the reversible transitions is higher in the films with lower $T_{Gr}$ and compensates for the drop in hardness that would arise from the increased rate of densification. On the other hand, the increase in stiffness of the network rings in films with higher $T_{Gr}$ is not sufficient to compensate for the increased rate of densification and we see a larger drop in their experimentally measured hardness.

## 4. Conclusion

We have used physical vapor deposition to access a range of different structures in a-SiO$_2$ and shown the relationships between structure and mechanical properties, which is supported by MD simulations. We have shown that there are nanovoids present in the films whose size and overall porosity increase with increasing growth temperature. This finding combined with the observed increase in density suggests that the a-SiO$_2$ network may become more compact with smaller rings as the growth temperature increases. We have also shown that the above changes in the structure of a-SiO$_2$ with increasing growth temperature result in higher elastic and shear moduli and higher hardness. These changes in mechanical properties have been attributed to the increase in density of the a-SiO$_2$ films. Furthermore, we have demonstrated that the vapor deposited a-SiO$_2$ films exhibit an anomalous increase in the elastic modulus with increasing indentation temperature similar to bulk a-SiO$_2$. However, the rate of change in the elastic modulus is found to depend on the growth temperature. We attribute this trend to the effect that the increase in density of the a-SiO$_2$ network has on the local structural transitions, which in turn are responsible for the anomalous increase in elastic and shear moduli with growth temperature.

In addition to providing fundamental insights on the structure-mechanical properties of a-SiO$_2$, the current study can aid in selecting growth parameters for the synthesis of a-SiO$_2$ films with tailored mechanical properties for specific applications.

**Acknowledgement**

This research was primarily supported by NSF through the University of Wisconsin Materials Research Science and Engineering Center (DMR-1720415).  T.H.M. and X.L. acknowledge support from the Office of Naval Research.

**Competing interests**

The authors declare no competing financial interests.

**Figures**

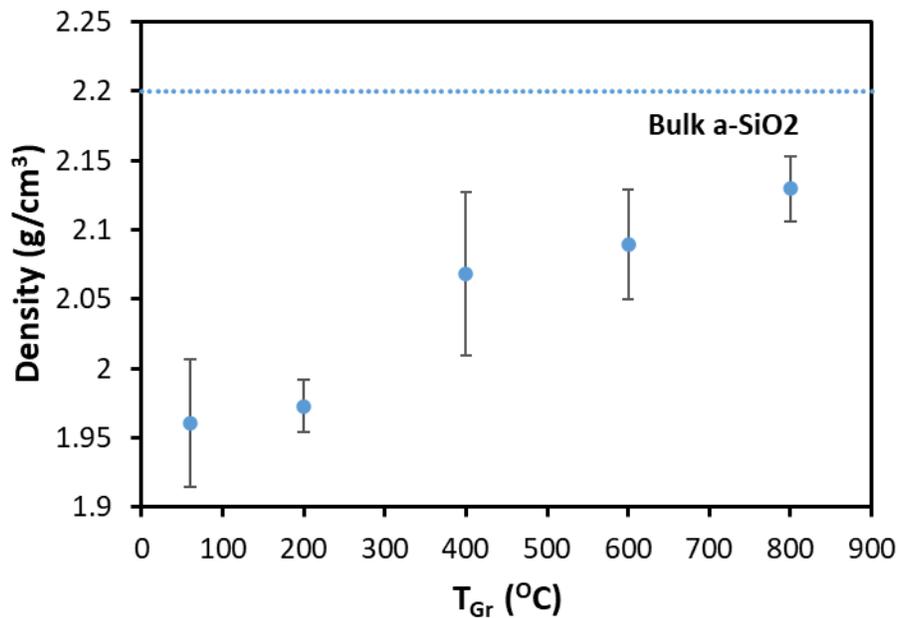

**Fig.1 Density of a-SiO$_2$ films determined using RBS in combination with profilometry. The density of the films increases with increasing growth temperature. Bulk a-SiO$_2$ density from Ref.[43]. Error bars represent standard errors calculated through error propagation from experimental errors in the RBS spectra and profilometry measurements.**

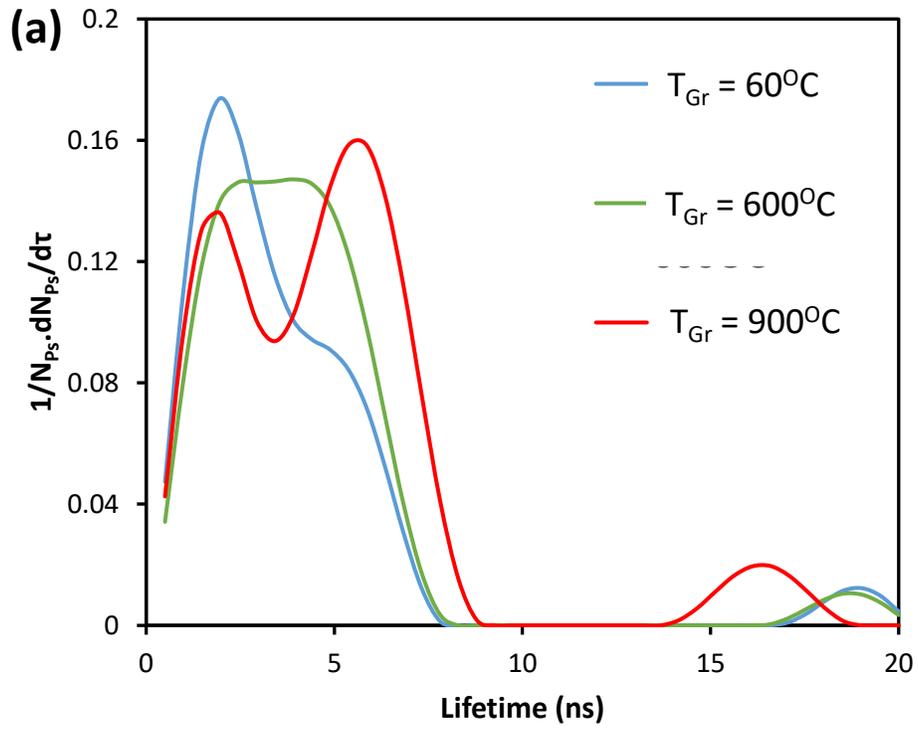

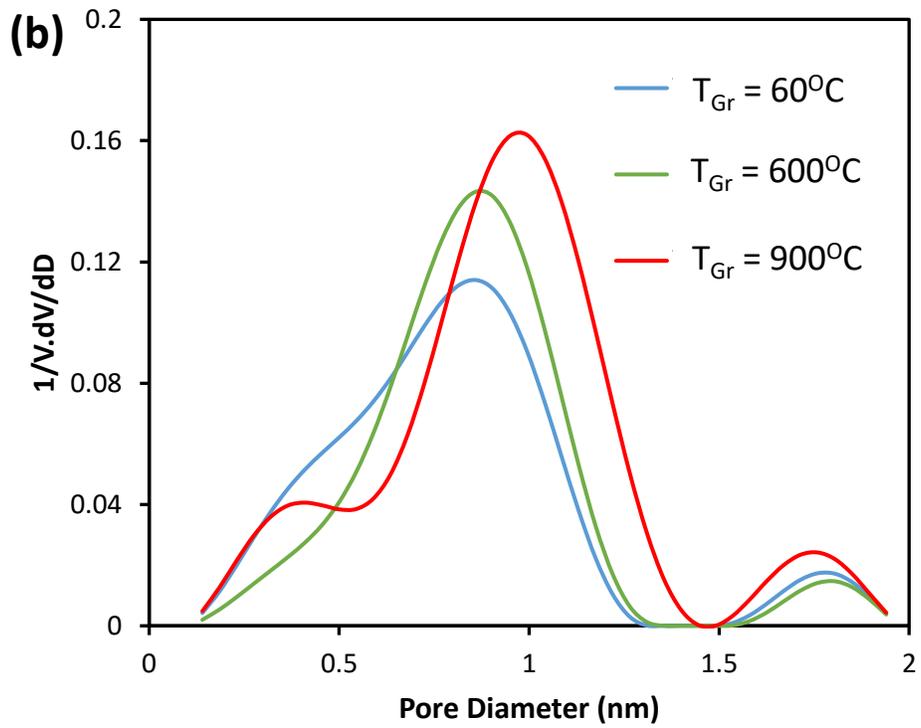

Fig.2 (a) Positronium lifetime distributions for a-SiO$_2$ films grown at different temperatures fitted with the program Contin. The total area under each distribution is determined by the relative total positronium formation fraction of each sample (900 °C highest, 60 °C lowest) as discussed in the text. (b) The relative pore volume distribution for a-SiO$_2$ films grown at different temperatures deduced by directly fitting each lifetime spectrum with the Contin program using a spherical pore model conversion of lifetime to pore diameter.

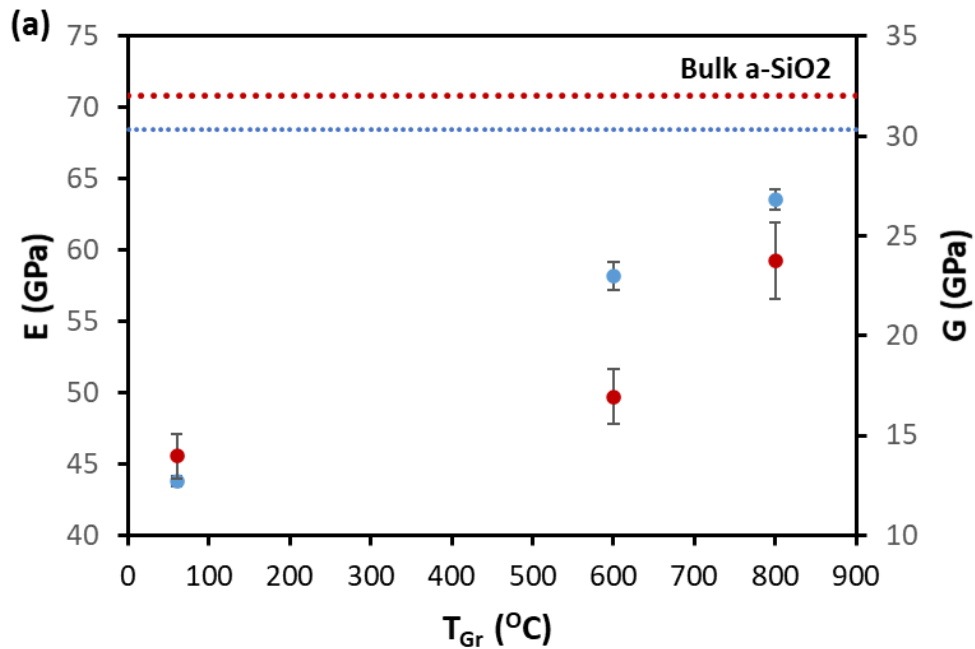

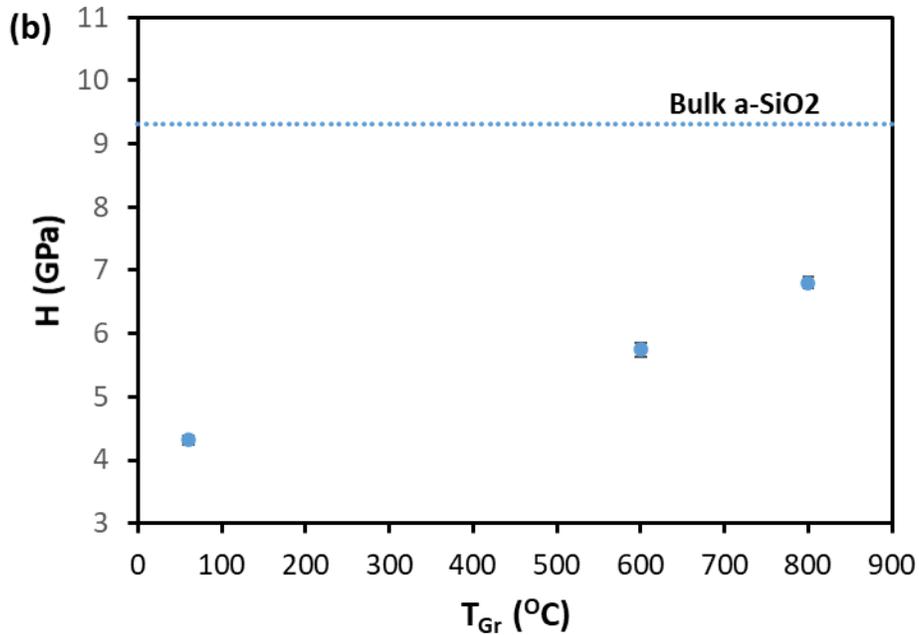

Fig.3 (a) Elastic modulus E (blue) and shear modulus G (red) and (b) Hardness H of a-SiO$_2$ films as a function of growth temperature. E, G, and H were measured experimentally using nanoindentation and DPOs. Dotted lines indicate elastic modulus and hardness of bulk a-SiO$_2$ measured from nanoindentation and shear modulus from Ref.[44]. Error bars in E and H correspond to the standard errors in the mean and are determined based on 15 indents on each sample. The error bars in G are standard errors determined by error propagation from the resonator frequency resolution measured by a lock-in amplifier and the error in thickness measured by profilometry.

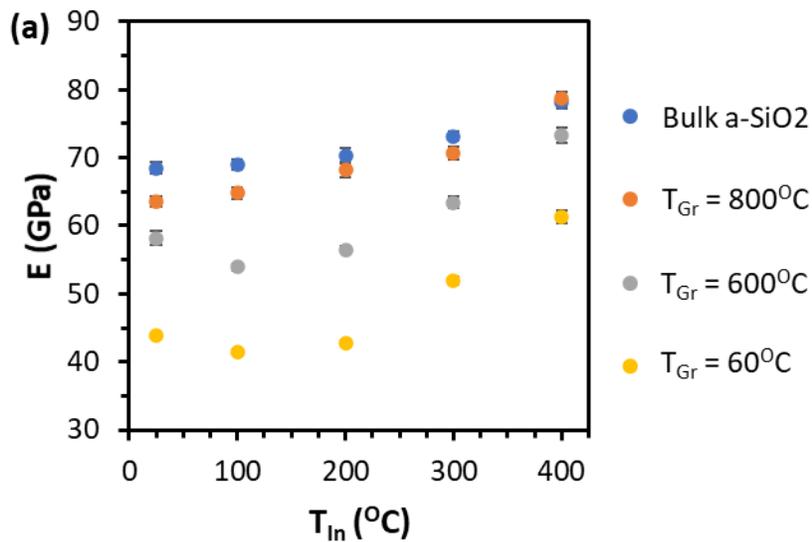

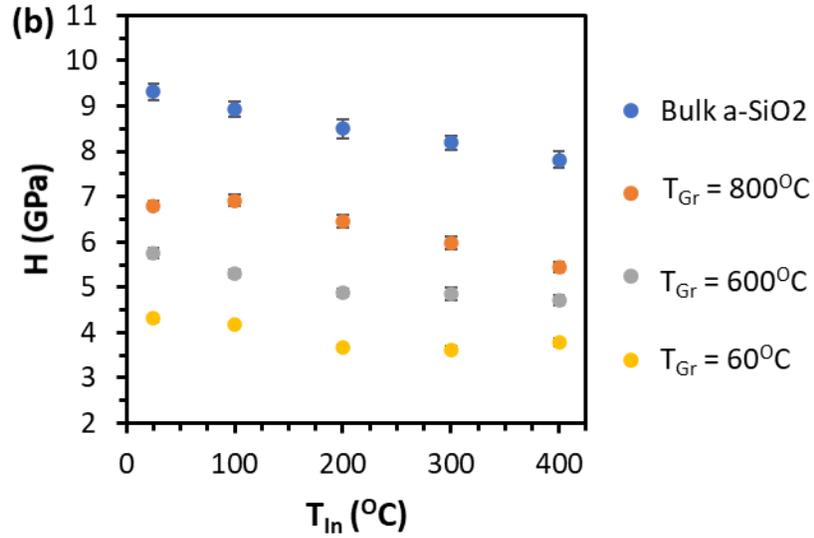

Fig.4 Experimentally measured change in the (a) elastic modulus and (b) hardness with indentation temperature $T_{In}$ for a-SiO$_2$ films grown at different temperatures ($T_{Gr}$) and a bulk a-SiO$_2$ sample. Error bars correspond to the standard errors in the mean and are based on 15 indents on each sample at each indentation temperature.

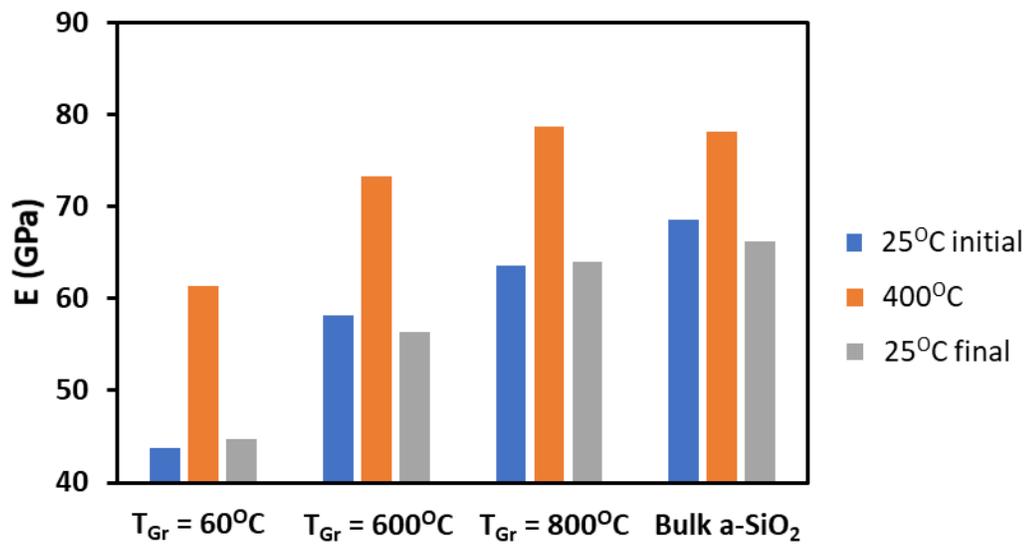

Fig. 5 Comparison of the elastic modulus before heating, after heating to 400 °C, and after cooling back down to room temperature.

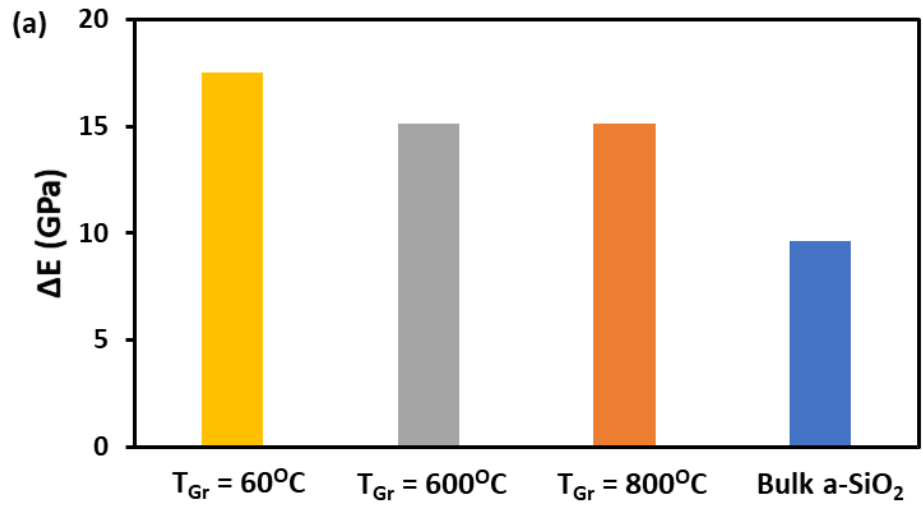

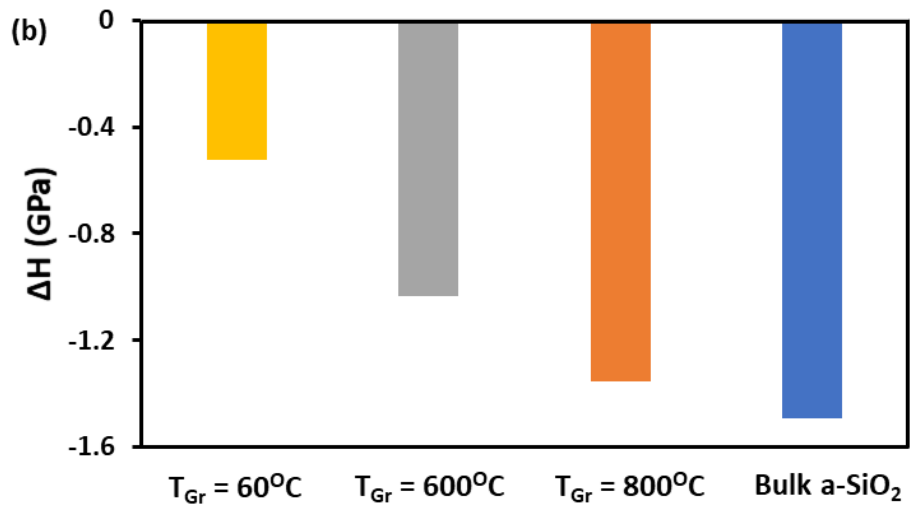

Fig.6 Experimentally measured changes in (a) elastic modulus ΔE and (b) Hardness ΔH when indentation temperature is increased from 25 °C to 400 °C

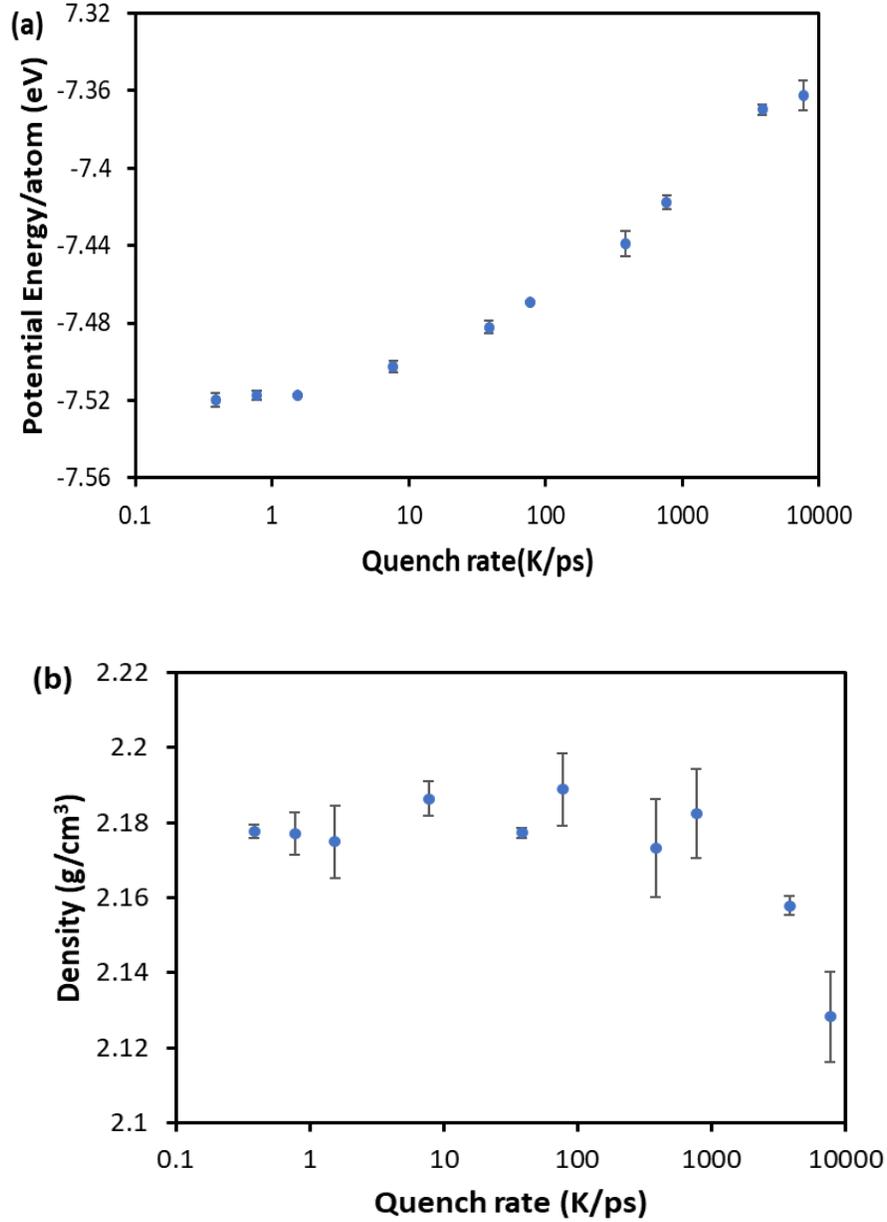

Fig. 7 (a) Potential energy per atom and (b) density of melt-quenched a-SiO$_2$ as a function of the quench rate from MD simulations. Error bars represent the standard deviation from the mean for three different initial configurations of the melt quenched structures.

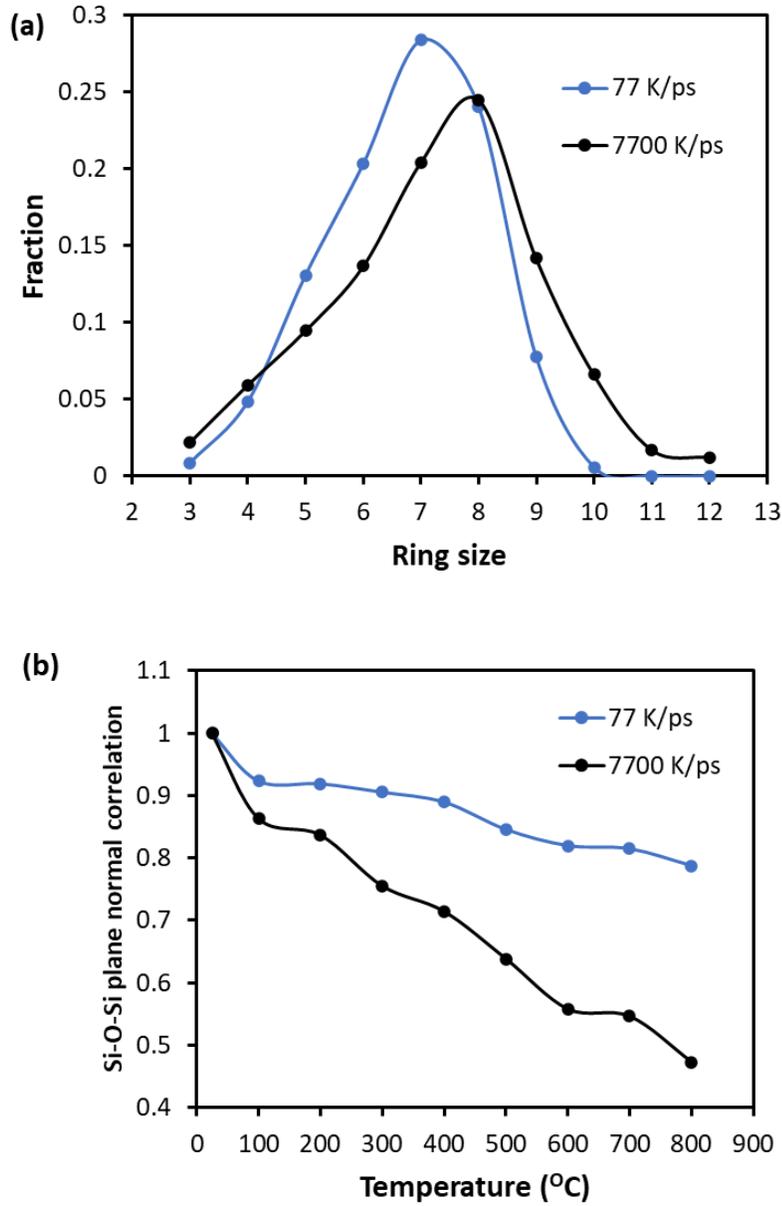

**Fig.8 (a)** Ring size distributions for melt-quenched a-SiO$_2$ structures from MD simulations, determined by R.I.N.G.S code [42]. Fraction of larger rings (> 9- members) increases as the cooling rate increases/density decreases. **(b)** Si-O-Si plane normal correlation as a function of temperatures for melt quenched structures with different cooling rates from MD simulations. Structure with a higher cooling rate/lower density shows a larger drop in the Si-O-Si plane normal correlation compared to lower cooling rate/higher density structure.